\documentclass[12pt]{article}

\usepackage{amsmath,amssymb,amsthm} 
\usepackage{appendix} 
\usepackage{bm} 
\usepackage{mathrsfs} 
\usepackage{multirow} 
\usepackage{float} 
\usepackage{bbm} 
\usepackage{slashed} 

\usepackage{latexsym}
\usepackage{amsmath}
\usepackage{amsfonts}
\usepackage{amssymb}
\usepackage{amscd}
\usepackage{amsthm}

\usepackage{bbm}
\usepackage{fancybox}
\usepackage{cite}
\usepackage{amsmath,amsfonts,amsbsy}
\usepackage{pstricks,pst-node}
\usepackage[small,bf,hang]{caption2}
\usepackage{slashed}
\usepackage{graphicx}
\usepackage{epsfig}
\usepackage{psfrag}
\usepackage{comment}
\usepackage[normalem]{ulem}
\usepackage{tikz}
\newtheorem{theorem}{Theorem}

\usepackage{float}
\usepackage{enumerate}
\psset{unit=1.3cm,linewidth=.5pt,radius=.2} 

\usepackage{multirow}           
\usepackage{float}             
\usepackage{lscape}             
\usepackage{bm}

\newcommand{\one}{\mathbbm{1}} 

\DeclareMathOperator{\tr}{tr} 
\newcommand{\ket}[1]{\left| #1 \right\rangle} 
\newcommand{\pd}{\partial} 
\renewcommand{\L}{\mathscr{L}} 
\newcommand{\adj}[2]{#1^{\dagger #2}} 
\newcommand{\comm}[2]{\left[ #1,#2\right]} 
\newcommand{\acomm}[2]{\left\{#1,#2\right\}} 
\newcommand{\osp}{\mathfrak{osp}} 
\newcommand{\ospr}{\mathfrak{osp}(4^*|4)} 
\newcommand{\so}{\mathfrak{so}} 
\newcommand{\usp}{\mathfrak{usp}} 
\newcommand{\su}{\mathfrak{su}} 
\newcommand{\kahler}{K\"{a}hler } 
\renewcommand{\a}{\alpha} 
\renewcommand{\b}{\beta}

\renewcommand{\d}{\delta}
\newcommand{\e}{\epsilon}

\newcommand{\m}{\mu}
\newcommand{\n}{\nu}
\renewcommand{\r}{\rho}
\newcommand{\s}{\sigma}

\newcommand{\D}{\Delta}

\newcommand{\be}{\begin{eqnarray}}
\newcommand{\ee}{\end{eqnarray}}
\newcommand{\nn}{\nonumber}

\newcommand{\beq}{\begin{equation}}
\newcommand{\eeq}{\end{equation}}
\newcommand{\bi}{\begin{itemize}}
\newcommand{\ei}{\end{itemize}}
\newcommand{\bt}{\begin{tabular}}
\newcommand{\et}{\end{tabular}}
\newcommand{\bc}{\begin{center}}
\newcommand{\ec}{\end{center}}

\newcommand{\baa}{\begin{equation}\begin{aligned}}
\newcommand{\eaa}{\end{aligned}\end{equation}}
\newcommand{\bea}{\begin{eqnarray}}
\newcommand{\eea}{\end{eqnarray}}
\newcommand{\ba}{\begin{array}}
\newcommand{\ea}{\end{array}}

\def\bbox{{\,\lower0.9pt\vbox{\hrule \hbox{\vrule height 0.2 cm
\hskip 0.2 cm \vrule height 0.2 cm}\hrule}\,}}
\newcommand{\dsl}{\pa \kern-0.5em /}

\def\tr{{\rm tr}}

\makeatletter \@addtoreset{equation}{section} \makeatother

\title{An Index for Superconformal Quantum Mechanics}

\author{Nick Dorey and Andrew Singleton\\
Department of Applied Mathematics and Theoretical Physics, \\
University of Cambridge, \\
Cambridge, CB3 OWA, UK \\
{\tt  n.dorey@damtp.cam.ac.uk, asingleton921@gmail.com} \\}

\begin{document}
\maketitle
\abstract{We study quantum mechanical systems with 
$\ospr$ superconformal symmetry. We classify unitary lowest-weight 
representations of this superconformal algebra and define an index
which receives contributions from short and semi-short multiplets only.  
We consider the example of a quantum 
mechanical $\sigma$-model with hyper-K\"{a}hler target 
$\mathcal{M}$ equipped with a triholomorphic homothety.  
The superconformal index coincides with the Witten index of a novel form of 
supersymmetric quantum mechanics for a particle moving on
$\mathcal{M}$ in a background magnetic field in which an 
unbroken $\mathfrak{su}(1|2)$ subalgebra 
of the superconformal algebra is linearly realised as a global symmetry.}

\maketitle


\section{Introduction}
Superconformal field theories (SCFTs) are of great interest, in part because
they provide the best understood examples of the 
AdS/CFT correspondence \cite{Maldacena:1997}. 
In these theories, the spectrum of local
operators naturally decomposes into 
irreducible representations of the
superconformal algebra. Amongst these the BPS
representations play a special role; they contain primary fields whose
dimensions saturate a lower bound and 
do not receive quantum corrections. The counting of such
states is complicated by the fact that BPS representations can combine
to form generic representations whose dimensions can then be
corrected. Although the spectrum of BPS states can change one can
define a {\em superconformal index} 
\cite{Kinney:Maldacena:Minwalla:Raju:2005}, \cite{Rom} which is invariant under 
variations of marginal couplings.       
\paragraph{}
Superconformal quantum mechanics provides a simpler setting where many
of the same phenomena occur. Such models are also of independent
interest as they 
arise in the DLCQ description of higher dimensional SCFTs and should
also play a role in the, so-far elusive, boundary
description of $AdS_{2}$. In this paper we will describe the BPS
sector of superconformal quantum mechanics and formulate an index
analogous to the superconformal index of 
\cite{Kinney:Maldacena:Minwalla:Raju:2005}. A more detailed 
discussion of the results presented here can be found in the PhD
thesis of the second-named author \cite{AS:thesis}. 
\paragraph{}
Although aspects of our 
construction should apply more generally, we will 
focus on a particular family of quantum-mechanical $\sigma$-models. 
For any Riemannian target $\mathcal{M}$ admitting a closed homothety, 
a bosonic $\sigma$-model has an $\mathfrak{so}(2,1)$ 
conformal symmetry \cite{Michelson:Strominger:1999}. 
If $\mathcal{M}$ is hyper-K\"{a}hler and the
homothety is triholomorphic, then the corresponding supersymmetric
$\sigma$-model has a superconformal symmetry isomorphic to the
simple Lie superalgebra $\osp(4^*|4)$ \cite{AS:2014}. 
This condition is satisfied by a large class of singular spaces, known
as hyper-K\"{a}hler cones\cite{dWRV}, which
arise as Higgs branches of supersymmetric gauge theories with eight
supercharges. 
\paragraph{}
In order to properly formulate supersymmetric quantum mechanics on a
hyper-K\"{a}hler cone, a resolution of its singularities is required. 
In \cite{DBG},
one of the authors with Barns-Graham propose a regulated definition
of the index, where the singular cone is replaced by its equivariant
symplectic resolution. Although, the resolution breaks superconformal 
invariance, 
a smaller algebra corresponding to the stabilizer of the 
BPS bound is preserved, allowing the definition of a regulated 
superconformal index on the resolved space. In \cite{DBG}, evidence is
presented that the index is independent of the choice of resolution
and includes information about the spectrum of superconformal
multiplets associated with the unresolved space. 
Via localisation, the resulting definition also yields a closed
formula for the index in many cases. In the final section of the paper
we will use the formula of \cite{DBG} to discuss some examples.   
\paragraph{}
This class of target spaces obeying the conditions for superconformal
invariance outlined above includes the moduli space of 
Yang-Mills instantons on $\mathbb{R}^{4}$. 
This example provides further support for the existence of  
$\ospr$-invariant superconformal quantum mechanics. In particular, 
such a model should provide a discrete lightcone description of
the $(2,0)$ theory in six dimensions \cite{Ah:1997},\cite{ABS:1997}. 
In this context,
$\osp(4^*|4)$ is part of the subalgebra of the $(2,0)$ superconformal
algebra which is preserved by compactification of a null direction.       
Thus we expect the model to have a discrete spectrum of lowest-weight unitary
representations of $\ospr$ arising from the branching of
six-dimensional $(2,0)$ multiplets onto the lightcone subalgebra.   
In this paper, we will consider the general properties of any such $\ospr$
invariant model. 
We classify the unitary irreducible representations of $\ospr$,
identifying various types of short and semi-short representations. We define a
superconformal index which by construction is invariant under the allowed 
recombinations of semi-short multiplets into generic ones.                 
\paragraph{}
The most general index for $\ospr$ consists of any way of counting 
(semi)short representations which is invariant under continuous 
supersymmetry-preserving deformations of the model of interest. In 
particular, representations which can pair into long representations 
must not contribute to this counting. These arguments lead to a basis of
`elementary indices' of which the superconformal index must be a linear 
combination. The superconformal index itself is defined analogously to 
the Witten index, as a supertrace, 
\begin{equation} \label{eqn:indexdef} I(t,y)=
\tr \left[ (-1)^F e^{-\b \mathcal{H}} t^{\mathbb{T}}
  y^{\mathbb{N}}\right]. 
\end{equation}
where the `Hamiltonian' $\mathcal{H}$ (with eigenvalue $E$) 
vanishes on states saturating a 
certain BPS bound.  This corresponds to 
a choice of supercharge $q$ and conjugate $s=\adj{q}{}$ with 
$\mathcal{H}=\acomm{q}{s}$. Here $t$ and $y$ are fugacities 
for the Cartan generators $\mathbb{T}$ and $\mathbb{N}$ of
the subalgebra $\su(2|1)\subset
\ospr$ which commutes with $\left\{q,s,\mathcal{H}\right\}$. 
The index can be further refined by including fugacities for charges
corresponding to the Cartan subagebra of the global symmetry group. 
As with the usual Witten index, $I(t,y)$ receives contributions 
only from states 
with $E=0$, and evaluating these leads to a general expression for the 
superconformal index as a sum of characters of $\su(2|1)$ multiplying the 
elementary indices. 
\paragraph{}
The problem of identifying and counting states with $E=0$ can also be recast as 
an ordinary supersymmetric quantum mechanics problem, albeit with somewhat 
exotic supersymmetry. In particular, starting from the original
$\sigma$-model, one may construct a Lagrangian whose Legendre
transform concides with the classical limit of the Hamiltonian,
$\mathcal{H}$. The ``little-group'' algebra $\su(2|1)$ is linearly realised as a
global symmetry of the Lagrangian. 
The resulting model can be interpreted in terms of 
a particle in a magnetic field proportional to a target space \kahler form.
The states which contribute to the superconformal index lie in the 
lowest Landau level and can be isolated in the usual way by taking a
strong field limit. 
In the context of superconformal mechanics the limit is purely kinematic
and corresponds to subjecting the model to a large conformal boost. 
\paragraph{}
In the final section of the paper, we consider some examples. In these
cases we compute the superconformal index exactly and extract
information about the corresponding spectrum of superconformal
multiplets. In particular we will see that the spectrum of certain
types of multiplets can be determined exactly. 
  
\section{Superconformal quantum mechanics}

We start from the standard supersymmetric 
quantum mechanical $\sigma$-model whose target space is a 
Riemannian manifold $(\mathcal{M},g)$ of dimension $n$. The bosonic degrees of
freedom corresponding to coordinates $X^{\mu}$ on $\mathcal{M}$ are
accompanied by complex conjugate fermionic degrees of freedom
$\psi^{\mu}$ and $\adj{\psi}{\m}=\adj{\left(\psi^{\m}\right)}{}$ as
Grassmann-odd sections of the cotangent bundle, 
\begin{equation} S=\int dt \, \frac{1}{2} g_{\m
    \n}(X)\dot{X}^{\m}\dot{X}^{\n}+ig_{\mu \nu}(X)\adj{\psi}{\m}
\frac{D}{Dt}\psi^{\n}-\frac{1}{4}R_{\m \n \r
  \s}(X)\adj{\psi}{\m}\adj{\psi}{\n}\psi^{\r}\psi^{\s}. 
\label{eqn:standardS} \end{equation} 
where we define the covariant derivative
\begin{equation} \frac{D}{Dt}\psi^{\m} = \nabla_{\dot{X}}
\psi^{\m}=\dot{\psi^{\m}}+\dot{X}^{\n}\Gamma_{\n\r}^{\m}\psi^{\r},
\label{eqn:fermiddt} \end{equation}
and $R_{\m\n\r\s}$ is the Riemann tensor on $\mathcal{M}$. The action
is invariant under $\mathcal{N}=(1,1)$ supersymmetry transformations, 
\begin{align} \begin{aligned}
    \d_{\e}X^{\m}&=-\adj{\e}{}\psi^{\m}+\e\adj{\psi}{\m} \\
    \d_{\e}\psi^{\m} &=
    i\dot{X}^{\m}\e-\Gamma^{\m}_{\n\r}\left(\d_{\e}X^{\n}\right)\psi^{\r} 
\end{aligned} \label{eqn:N11SUSYtransformations} 
\end{align}
\paragraph{}
The phase space is parametrised by the coordinates $X^{\mu}$ and 
their canonically conjugate momenta $P_{\m} = {\pd \L}/{\pd
  \dot{X}^{\m}}$ together with the fermionic variables
$\psi^{\m}$, $\adj{\psi}{\m}$. It is often convenient 
to work with the covariant momentum, $\Pi_{\m} = g_{\m
  \n}\dot{X}^{\n}$ which transforms as a tangent vector on the
manifold. 
Quantization proceeds by imposing the
(anti-)commutation relations, 
\begin{eqnarray} 
\left[X^{\mu}, \Pi_{\mu}\right] = i\delta^{\mu}_{\nu} & \qquad{} &
\left\{ \psi^{\m}, \adj{\psi}{\n} \right\} = g^{\mu\nu} \nonumber 
\end{eqnarray}
As the covariant momentum is not canonical we have further
non-vanishing commutators, 
\begin{equation}
  \comm{\Pi_{\m}}{\psi^{\n}}=i\Gamma^{\n}_{\m\r}\psi^{\r}, \qquad
  \comm{\Pi_{\m}}{\adj{\psi}{\n}}=i\Gamma^{\n}_{\m\r}\adj{\psi}{\r},
  \qquad 
\comm{\Pi_{\m}}{\Pi_{\n}} = -R_{\r \s \m
  \n}\adj{\psi}{\r}\psi^{\s}.\label{eqn:basiccovariantcomm} 
\end{equation} 
We choose a fermionic vacuum $|0\rangle$ annihilated by all the fermions 
$\psi^{\m}$, $\m=1,\ldots, n$. The resulting Hilbert space is
naturally organised by the number of fermions excited. A state with
$p$ fermions, 
\begin{equation} \ket{\a}=\frac{1}{p!}\a_{\m_1\dots\m_p}(X)
\adj{\psi}{\m_1}\dots\adj{\psi}{\m_p}\ket{0}, 
\label{eqn:genericstate} \end{equation} 
is naturally identified with  the following differential form of degree $p$ on 
$\mathcal{M}$,
\[ \alpha = \frac{1}{p!}\alpha_{\m_1\dots\m_p}(X)dX^{\m_1}
\wedge\dots\wedge dX^{\m_p}.\]
The inner product between states correspond
to the standard $L^{2}$ inner product of forms, \begin{equation} 
\left\langle\alpha,\beta\right\rangle = \int_M \alpha \wedge
*\bar{\beta} = \frac{1}{p!}\int d^nX \sqrt{ g}
\,\a_{\m_1\dots\m_p}\bar{\b}^{\m_1\dots\m_p}, 
\label{eqn:L2product} \end{equation}
With this identification the Hamiltonian coincides with the Laplacian
acting on forms. Its explicit form, up to additional terms arising
from operator reordering, is given as,  
\begin{eqnarray}
\mathbb{H} & = &   g^{\m \n}\Pi_{\m}\Pi_{\n} + 
\frac{1}{2}R_{\m \n \r \s}\adj{\psi}{\m}\adj{\psi}{\n}\psi^{\r}\psi^{\s} 
\nonumber 
\end{eqnarray}
The supersymmetry transformations given above are generated by
supercharges, 
\begin{eqnarray}
\mathbb{Q} = i\adj{\psi}{\m}\Pi_{\m}, & \qquad{} \mathbb{Q}^{\dagger}= -i 
\Pi^{\dagger}_{\m}\psi^{\m} \nonumber 
\end{eqnarray} 
obeying the algebra, 
\begin{eqnarray}
\left\{\mathbb{Q}, \mathbb{Q}^{\dagger} \right\} & = &  \mathbb{H} 
\label{susy1}
\end{eqnarray} 
The supersymmetry algebra admits a $U(1)$ R-symmetry generated by, 
\begin{eqnarray}
\mathbb{J} & = & \frac{1}{2}\left(g_{\m
    \n}\adj{\psi}{\m}\psi^{\n}-\frac{n}{2}\right) =
\frac{1}{2}\left(p-\frac{n}{2}\right) 
\label{R}
\end{eqnarray}

So far our discussion is generic to any Riemannian target manifold. In
the special case that the $\mathcal{M}$ is a complex K\"{a}hler
manifold with complex structure $I^{\mu}_{\nu}$ we can define 
additional supercharges\cite{AlvarezGaume:Freedman:1981}, 
\begin{eqnarray}
\mathbb{Q}^{(I)} = i\adj{\psi}{\m}I_{\mu}^{\nu}\Pi_{\n}, & \qquad{} 
\mathbb{Q}^{(I)\dagger}= -i 
\Pi^{\dagger}_{\m}I^{\mu}_{\nu}\psi^{\n} \nonumber 
\end{eqnarray} 
Further, if the manifold is hyper-K\"{a}hler, with three linearly
independent complex structures $(I^{a})^{\m}_{\n}$, $a=1,2,3$ obeying the
$\mathfrak{su}(2)$ algebra, 
\begin{equation} I^aI^b=-\delta^{ab}\one + \e^{abc}I^c
\label{eqn:quaternionalgebra}\end{equation}
In this case we also have 
additional supercharges 
$\mathbb{Q}^{(a)}=i\adj{\psi}{\m}(I^{a})_{\mu}^{\nu}\Pi_{\n}$ and 
$\mathbb{Q}^{(a)\dagger}$ generating the $\mathcal{N}=(4,4)$
supersymmetry algebra where (\ref{susy1}) is supplemented by,  
\begin{eqnarray}
\left\{\mathbb{Q}^{a}, \mathbb{Q}^{b\dagger} \right\} & = &  
\delta^{ab}\mathbb{H} 
\nn 
\end{eqnarray} 
The $\mathcal{N}=(4,4)$ supersymmetry algebra admits an
$\mathfrak{so}(5)\simeq\mathfrak{sp}(4)$ R-symmetry
\cite{Verbitsky:1990} under which the complex
supercharges transform in the four dimensional spinor representation
denoted ${\bf 4}$.   
\paragraph{}
In this paper we will be interested in supersymmetric $\sigma$-models
which also admit a conformal symmetry and thus, in particular, we will
consider models with scale invariance. This requires 
\cite{Michelson:Strominger:1999} the target 
manifold $\mathcal{M}$ to admit a homothetic Killing vector field
${D}$ obeying, 
\be 
\mathcal{L}_{{D}}\, g & = & 2g \label{hom1}
\ee
where $\mathcal{L}_{D}$ denotes the corresponding Lie
derivative. Further, we say that ${D}$ is a {\em closed
  homothety} if there exists a real scalar function $K$ on
$\mathcal{M}$ such that, 
\begin{eqnarray} 
\mathcal{L}_{{D}}\, {K} =  2{K} & \qquad{} & D_{\mu}=\partial_{\mu}K 
\label{chom}
\end{eqnarray}
In this case we can define a {\em dilation operator} for the quantum
mechanics defined by the action (\ref{eqn:standardS}) as 
$\mathbb{D}=D^{\mu}\Pi_{\mu}-in/2$ and a {\em Special conformal
  generator} $\mathbb{K}=K$. By virtue of (\ref{hom1}) and 
(\ref{chom}) the three generators $\{\mathbb{D}, \mathbb{H},
\mathbb{K}\}$ yield an
$\mathfrak{so}(2,1)\simeq \mathfrak{su}(1,1)$ conformal symmetry, 
\begin{eqnarray}
\left[\mathbb{D},\mathbb{H}\right]= 2i\mathbb{H}, &\qquad{}  
\left[\mathbb{D},\mathbb{K}\right]= -2i\mathbb{K},
\qquad{} &    \left[\mathbb{H},\mathbb{K}\right]= -i\mathbb{D} 
\nn 
\end{eqnarray}
For the generic Riemannian target $\mathcal{M}$, the above conformal
algebra together with the $\mathcal{N}=(1,1)$ supersymmetry algebra
close onto an $\mathfrak{su}(1,1|1)$ superconformal symmetry algebra 
\cite{Michelson:Strominger:1999} with
maximal bosonic subalgebra
$\mathfrak{su}(1,1)\oplus\mathfrak{u}(1)$. The $\mathfrak{u}(1)$
factor is generated by the R-charge (\ref{R}) defined above. In
addition to the Poincare supercharges $\mathbb{Q}$ and
$\mathbb{Q}^{\dagger}$, the fermionic generators include the
superconformal charges, 
\begin{eqnarray}
\mathbb{S}=-i\left[\mathbb{K},\mathbb{Q}\right],  & \qquad{}\qquad{} & 
 \mathbb{S}^{\dagger}=-i\left[\mathbb{K},\mathbb{Q}^{\dagger}\right]  
\nn 
\end{eqnarray}
Thus $\mathbb{Q}$ and $\mathbb{S}$ are two components of a doublet of 
$\mathfrak{su}(1,1)$. 
\paragraph{}
Further enhancements of the superconformal symmetry can occur in the
cases of complex target manifolds discussed
above if the homothety $D$ is holomorphic with respect to one or more
complex structures: $\mathcal{L}_{D}I=0$. We will be interested in the 
maximal superconformal symmetry
which occurs when $\mathcal{M}$ is a hyper-K\"{a}hler manifold
admitting a tri-holomorphic closed homothety \cite{AS:2014}. 
In other words, in addition to the conditions discussed above, 
the homothety $D$ is  
holomorphic with respect to each of the linearly independent complex 
structures on $\mathcal{M}$: $\mathcal{L}_{D}I^{a}=0$ for $a=1,2,3$. 
In this case we find a superconformal algebra isomorphic to the simple
Lie superalgebra $\osp(4^*|4)$ \cite{Kac:1977}, \cite{Parker:1980}. 
The bosonic subalgebra,    
\[ \mathfrak{g}_B = \so(2,1)\oplus \su(2) \oplus \so(5)\]
contains the the conformal algebra together with the $\so(5)$ R
symmetry of the $\mathcal{N}=(4,4)$ supersymmetry algebra which acts
only on the $\sigma$-model fermions. The additional $\mathfrak{su}(2)$ 
acts only on the bosonic coordinates. The four Poincare supercharges 
$\{\mathbb{Q},\mathbb{Q}^{a}\}$, for $a=1,2,3$ together with the 
corresponding superconformal charges tranform in the
$(\mathbf{2},\mathbf{2},\mathbf{4})$ of $\mathfrak{g}_{B}$.  
\paragraph{}
To study the spectrum of the dilatation operator it is convenient to
perform a change of basis in $\mathfrak{su}(1,1)$ as, 
\begin{equation} X \mapsto e^{-\m \mathbb{K}}
  e^{\frac{1}{2}\m^{-1}\mathbb{H}}Xe^{-\frac{1}{2}\m^{-1}\mathbb{H}}e^{\m
    \mathbb{K}} := 
M^{-1}XM
  \qquad \forall X \in \so(2,1) \label{eqn:so21basischange} \end{equation}
with $\m \in (0,\infty)$, under which
\begin{align} \label{eqn:DHKchange} \begin{aligned} i\mathbb{D}
    &\mapsto 
\mathbb{L}_0 =
    \m^{-1}\left(\mathbb{H}+\m^2\mathbb{K}\right) \\ \mathbb{H} 
&\mapsto 2\m \mathbb{L}_- =
    \m\left(\m^{-1}\mathbb{H}-\m \mathbb{K} - i\mathbb{D}\right) \\ 
\mathbb{K} &\mapsto
    -\frac{1}{2\m}\mathbb{L}_+=-\frac{1}{4\m}\left(\m^{-1}\mathbb{H}-\m 
\mathbb{K} +i \mathbb{D}\right). \end{aligned} 
\end{align}
These satisfy
\[ \adj{\mathbb{L}_0}{}=\mathbb{L}_0, 
\qquad \adj{\mathbb{L}_+}{}=\mathbb{L}_-, \qquad \comm{\mathbb{L}_0}{
\mathbb{L}_{\pm}}=2\mathbb{L}_{\pm}, \qquad 
\comm{\mathbb{L}_+}{\mathbb{L}_-}=-\mathbb{L}_0, \]
As the supercharges $\mathbb{S}$ and $\mathbb{Q}$ form a
doublet of $\mathfrak{su}(1,1)$, we must also perform the same
rotation on the supercharges forming linear combinations 
which are eigenstates of $\mathbb{L}_{0}$. 
\paragraph{}
For any value of $\mu$, $\mathbb{L}_{0}$ is isospectral to 
$\mathbb{D}$. In the flat space example it is easy to see that
$\mathbb{L}_{0}$ has a discrete spectrum due to the presence of the
harmonic potential provided by $\mathbb{K}$. We expect to find the same
qualitative behaviour in the general case, and expect a discrete
spectrum for $\mathbb{L}_{0}$ and therefore also for $\mathbb{D}$. 
Of course this is quite different from ordinary quantum mechanics on a 
non-compact space which has a continuous spectrum of scattering states. 
\paragraph{}
So far we have seen that quantum mechanics on a hyper-K\"{a}hler cone
exhibits an $\ospr$ superconformal symmetry. The main hypothesis of 
\cite{DBG} is that associated to each such cone, there is a discrete
spectrum of unitary representations of this algebra. In the next
section we will examine the general features of such a theory.

\section{The superconformal index}
\paragraph{}
In each representation of $\ospr$ basis states are labelled by the 
eigenvalues of the Cartan generators of the bosonic subalgebra,    
\[ \mathfrak{g}_B = \so(2,1)\oplus \su(2) \oplus \usp(4)\]
We choose Cartan generators
$\mathbb{J}_{3}$ for $\mathfrak{su}(2)$ and $\mathbb{M}$, $\mathbb{N}$
for $\mathfrak{usp}(4)\simeq \mathfrak{so}(5)$. 
We will work with lowest-weight
representations and assume the existence of a primary state
$\ket{\D,j,m,n}$ with, 
\begin{align} \nonumber \begin{aligned}  
\mathbb{L}_{0}\ket{\D,j,m,n}&=\D\ket{\D,j,m,n} & {\mathbb J}_3\ket{\D,j,m,n} &= 
-j\ket{\D,j,m,n} \\ {\mathbb M}\ket{\D,j,m,n}&=-m\ket{\D,j,m,n} &
\qquad {\mathbb N}
\ket{\D,j,m,n}&=-n\ket{\D,j,m,n}.
\end{aligned} \end{align}
which is annihilated by all the lowering operators of the superconformal algebra, 
\begin{eqnarray}
\mathfrak{g}^-\ket{\D,j,m,n}&= & 0 
\end{eqnarray}
Here $2j\in \mathbb{N}$ and $m,n\in \mathbb{N}$ with $m\geq n$ label
the $\mathfrak{su}(2)$ and 
$\mathfrak{usp}(4)$ R-symmetry representations of the superconformal primary state, while
$\D\geq 0$ is its dimension. The quantum numbers of this state 
are chosen such that the module generated by the action of $\mathfrak{g}_B$ is 
unitary. With such a choice, one can show that the analysis of 
reducibility and unitarity for the resulting representation of 
$\ospr$ reduces to calculating the norms of states of the schematic form
\begin{equation} \ket{\mathbf{n},\D,j,m,n} := {Q}_1^{n_1}\dots 
{Q}_8^{n_8}\ket{\D,j,m,n}, 
\qquad n_i\in \left\{0,1\right\}.\label{eqn:FermiVerma} \end{equation}
Where ${Q}_{\alpha}$ $\alpha=1,2,\ldots,8$ denote the eight
real components of the complex supercharges $\mathbb{Q}$, $\mathbb{Q}^{a}$ defined
above. If there are no states of negative norm then the representation
can be made unitary and irreducible by quotienting out 
any states of zero norm. This construction makes the 
relationship between shortening of 
representations and the presence of BPS states, which are annihilated
by one or more supercharge, manifest.
\paragraph{}
By calculating these norms for the case of one or two supercharges 
acting, we can easily obtain some necessary conditions for unitarity. 
Proving sufficiency is somewhat harder work, requiring a link between 
the presence of zero norm states and the atypicality conditions of 
\cite{Kac:1977,Kac:1978}. The details of this construction are
described in \cite{AS:thesis}. 
The upshot is the following classification:
\begin{theorem} 
Unitary, irreducible, lowest weight 
representations of $\ospr$ are obtained from the Verma module generated 
by the action of $\ospr$ on $\ket{\D,j,m,n}$, by quotienting out null 
states. They come in the following types:
\begin{itemize} \item Generic `long' representations $L(\D,j,m,n)$ 
with $\D>2(j+m+1)$.
\item `Semishort' representations $SS(j,m,n)$ with
  $\D=\D_{SS}=2(j+m+1)$, $m\geq n$. 
\item `Short' representations $S(m,n)$ with $\D=2m$ and $j=0$, $m\geq n$. These 
split into 1/2-BPS representations with $m=n$ and 1/4-BPS otherwise.
\end{itemize}
\nonumber{}
\end{theorem}
Long representations contain no null states of the form 
(\ref{eqn:FermiVerma}), and consequently have no BPS states. The 
lowest weight state of a short representation is 1/2-BPS for $m=n$ 
and 1/4-BPS otherwise, while semishort representations contain BPS 
states at higher levels.
\paragraph{}
A key feature of (semi-)short representations and their BPS states is 
that, since their dimension is tied to their R-charges, it cannot 
vary continuously with parameters of a theory, and in particular is 
protected from quantum corrections. In terms of the representations 
above, one simply observes that (semi)short representations contain 
fewer states than long ones, so cannot continuously change their 
type. This argument is not quite correct, since a long representation 
of dimension $\D=\D_{SS}+\e:0<\e<<1$ can continuously lower 
$\e\rightarrow 0$. When this occurs, null states appear and the 
representation splits into a semishort representation with manifestly 
positive norm, and further (semi)short representations with zero norm. 
Dually, certain (semi)short representations can pair into a long one 
and move away from the unitarity bound. There are two ways this can 
occur:\footnote{Strictly speaking these decompositions are conjectural, 
as we do not at present have a proof that the null representations are 
irreducible. This would require either a computation of the characters, 
or a more careful analysis of representation structure as in 
\cite{Dolan:Osborne:2002}.}
\begin{align} \label{eqn:reppairs} \begin{aligned} L(\D_{SS}+\e,j,m,n)
&\rightarrow SS(j,m,n) \oplus SS(j-\frac{1}{2},m+1,n)&&(j>0) \\ 
L(\D_{SS}+\e,0,m,n) &\rightarrow SS(0,m,n)\oplus S(m+2,n) && (j=0). 
\end{aligned} \end{align}
Notice in particular that short representations with $m-n\leq 1$, in 
particular 1/2-BPS representations, cannot pair up so are absolutely 
protected.
\paragraph{}
By an {\em index} for $\ospr$ we mean any counting of (semi)short 
representations which is invariant under continuous changes of 
parameters. In particular, it must evaluate to zero on any 
combination of (semi)short representations which can pair into a 
long representation. That is, if $\mathscr{R}$ labels the set of 
possible (semi)short representations, we can define an index as
\[ I_{\a} = \sum_{R\in \mathscr{R}} \a(R)N(R),\]
where $N(R)$ is the number of representations of a given type $R$ 
present and $\a(R)$ are coefficients chosen such that $I_{\a}$ is 
an index. The decompositions (\ref{eqn:reppairs}) imply that these 
coefficients satisfy
\begin{align} \label{eqn:alphacontraints} \begin{aligned}
0 &= \a(SS(j,m,n) + \a(SS(j-\frac{1}{2},m+1,n)) &&(m\geq n\geq0,
~j>0) \\ 0 &= \a(SS(0,m,n) + \a(S(m+2,n))&& (m\geq n\geq 0).
\end{aligned} \end{align}
Solving these constraints gives a basis of elementary indices
\begin{equation} \label{eqn:indexbasis} I^{r,s} = \left\{ 
\begin{array}{ll} N(S(r,s)) &(r-s\leq 1) \\ N(S(r,s)) + 
\sum_{t=0}^{r-s-2} (-1)^{t+1} N(SS(\frac{t}{2},r-2-t,s))\quad & 
(r-s>1).\end{array} \right. \end{equation}
Any index for $\ospr$ must be a linear combination of these quantities. 

\paragraph{}
The superconformal index, originally defined for four-dimensional 
field theory in \cite{Kinney:Maldacena:Minwalla:Raju:2005} and 
extended to three, five and six dimensions in \cite{Bhattacharyaetal:2008} 
is formulated by selecting a supercharge $q$ (with Hermitian conjugate 
$s=q^{\dagger}$) which only vanishes on particular states
in (semi-)short multiplets. These states correspond to the
supersymmetric ground states of the corresponding ``Hamiltonian'' 
$\mathcal{H}=\{q,s\}$ with zero energy. 
The superconformal index is then defined as the corresponding 
Witten index of the form $\mathcal{I}={\rm
  Tr}[(-1)^{F}\exp(-\beta\mathcal{H})\ldots]$ where the dots denote
the possiblity of inserting other operators which commute with the
supercharges and $\mathcal{H}$. By standard arguments this quantity is
independent of $\beta$ (and of all other smooth deformations which
preserve supersymmetry) provided the spectrum of $\mathcal{H}$ is
discrete. The result is a quantity which recieves contributions only
from certain representatives of the (semi-)short multiplets.    
\paragraph{}
Here we will pick conjugate supercharges $q$
and $s=q^{\dagger}$ such that, 
\begin{eqnarray}
\{q,s\} \,=\, \mathcal{H} & =&
\mathbb{L}_{0}+2\mathbb{J}_{3}+2\mathbb{M} 
\nonumber 
\end{eqnarray}
In the context of the $\sigma$-model defined above, this implies picking a preferred complex structure
$\mathcal{I}^{\mu}_{\nu}$ on the target space, together with the corresponding holomorphic and
antiholomorphic projectors, 
\begin{eqnarray}
\left(\mathcal{P}_{\pm}\right)^{\mu}_{\nu} & = & \frac{1}{2}
\left(\delta^{\mu}_{\nu} \pm i \mathcal{I}^{\mu}_{\nu}\right)
\nn 
\end{eqnarray}
In terms of which, 
\begin{eqnarray}
\mathbb{N} & = & \left(\mathcal{P}_{-}\right)^{\mu}_{\nu}
\adj{\psi}{\n}\psi_{\m} \,-\, d 
\nn \\ 
\mathbb{M} & = & \left(\mathcal{P}_{+}\right)^{\mu}_{\nu}
\adj{\psi}{\n}\psi_{\m} \,-\, d \nn \\ 
\mathbb{J}_{3} & = & -\frac{1}{2}\mathcal{I}^{\m}_{\n}D^{\n}\Pi_{\m}
\nn 
\end{eqnarray}
where $d$ is the quaternionic dimension of the target manifold
$\mathcal{M}$. With this choice of complex structure, differential
forms on $\mathcal{M}$ are graded according to their holomorphic and
anti-holomorphic degrees $p$ and $q$, which are the eigenvalues of the
operators $\mathbb{N}+d$ and $\mathbb{M}+d$ respectively. This
identification implies bounds on the values of eigenvalues $d-n$ and
$d-m$ of these operators. In particular, as the 
(anti-)holomorphic
degrees of forms are bounded by the complex dimension of the 
target space, the integers $m$ and $n$ must lie in the 
interval  $[-d,+d]$.   
\paragraph{}
The choice of supercharges described above breaks the full
superconformal algebra down to the subalgebra 
$\mathfrak{su}(1|2)\subset \osp(4^*|4)$ spanned by generators
(anti-)commuting with both $q$ and $s$. The bosonic subalgebra, 
\begin{eqnarray}
\mathfrak{u}(1)\oplus \mathfrak{su}(2)& \subset & \mathfrak{su}(1|2)
\end{eqnarray}
of this ``little group'',
has Cartan subalgebra generated by $\mathbb{T}=-
(\mathbb{M}+ 2\mathbb{J}_{3})$ and $\mathbb{N}$. Now suppose that, in
addition to these generators we have some additional global symmetry
algebra of rank $r$ 
with commuting generators $J_{i}$, $i=1,2,\ldots,r$. The most general
superconformal index then takes the form, 
\begin{eqnarray}
\mathcal{I}\left(t,y,\{\mu_{i}\}\right) & = & 
{\rm Tr}\left[ (-1)^{F}e^{-\beta \mathcal{H}}\, 
t^{\mathbb{T}}y^{\mathbb{N}}e^{\sum_{i}\mu_{i}J_{i}}\right] 
\nn 
\end{eqnarray}
As the spectrum of $\mathbb{L}_{0}$ is discrete, the resulting index
is independent of $\beta$ by construction. Because
the little group generators commute with $q$ and $s$, they have a well
defined action on the space of states annihilated by $\mathcal{H}$ and
the index can therefore be decomposed in terms of the corresponding
characters. Thus, 
\begin{eqnarray}
\mathcal{I}\left(t,y,\{\mu_{i}\}\right) & = & \sum_{R}\, C_{R}
\left(\{\mu_{i}\}\right)\,\,\chi_{R}\left(t,y\right) 
\nonumber 
\end{eqnarray}
The sum is over irreducible representations $R$ of
$\mathfrak{su}(2|1)$ with character $\chi_{R}$. The index can
also be decomposed in characters of the global symmetry. 
Each short or semi-short representation of the
full superconformal algebra contains states with $\mathcal{H}=0$ which
contribute to the index. In each case the resulting states form a
module for a representation of the little group $\mathfrak{su}(2|1)$ 
and the contribution to the index is precisely the character of 
this representation.    
In order to decode the information contained in the index 
we must determine the $\mathfrak{su}(2|1)$ character
corresponding to each short or semi-short representation of
$\osp(4^{*}|4)$. The results are as follows: for $1/2$-BPS 
short representation $S(m,m)$ with $m\geq0$ and the  $1/4$-BPS 
short representation $S(m,n)$ with $m>n\geq 0$ we find characters, 
\begin{eqnarray}
\mathcal{I}_{m,m}\left(t,y\right) & = & t^{m}\left[\chi_{m}(y)-t
  \chi_{m-1}(y)\right] \nn \\ 
{\mathcal I}_{m,n}\left(t,y\right) & = & t^{m}\left[(1+t^{2})
\chi_{n}(y)-t\left(\chi_{n+1}(y)+\chi_{n-1}(y)\right)\right] \nn 
\end{eqnarray} 
where $\chi_{n}(y)$ is the character of the spin $n/2$ representation
of $\mathfrak{su}(2)$; 
\begin{eqnarray}
\chi_{n}(y) & = &  y^{n}+y^{n-2}+\ldots +y^{-n} \nn \end{eqnarray}
Thus $\chi_{0}(y)=1$ and we adopt the convention that
$\chi_{-1}(y)=0$. Similarly the semi-short representation $SS(j,m,n)$
yields the character, 
\begin{eqnarray}
\mathcal{I}_{j,m,n}\left(t,y\right) & = &  t^{2j+2}\mathcal{I}_{m,n}
\left(t,y\right)=t^{m+2j+2}\left[(1+t^{2})
\chi_{n}(y)-t\left(\chi_{n+1}(y)+\chi_{n-1}(y)\right)\right] \nn 
\end{eqnarray} 
Both types, of short multiplet have a lowest weight state with $E=0$ 
which becomes the lowest weight of the corresponding
$\mathfrak{su}(2|1)$ representation. For semi-short multiplets, states
with $E=0$ appear instead at the first excited level of the
$\ospr$ representation. In particular the corresponding lowest weight of
$\mathfrak{su}(2|1)$ is a level one state of the $\ospr$
representation with respect to the standard triangular decomposition
of the Lie superalgebra. 
\paragraph{}
We can now express the index in terms of the the numbers $N^{(m,n)}=N(S(m,n))$,
and $N^{(j,m,n)}=N(SS(j,m,n))$ 
of each type of representation present in the spectrum. In the case of
a $\sigma$-model, the multiplets appearing are subject to the
usual geometrical contraint on the (anti-)holomorphic degrees of forms on
the target space. Evaluating the unrefined
index with $\mu_{i}=0$ on a generic spectrum yields, 
\begin{eqnarray}
\mathcal{I}(t,y) & = & \sum_{m=0}^{d} N^{(m,m)} 
\mathcal{I}_{m,m}\left(t,y\right)\,-\, \sum_{m=1}^{d} N^{(m,m-1)} 
\mathcal{I}_{m,m-1}\left(t,y\right) \nonumber \\ 
& & \,+\,\sum_{m=n+2}^{\infty} \sum_{n=0}^{d-1} (-1)^{m-n}\, 
{\tilde N}^{(m,n)}\mathcal{I}_{m,n}\left(t,y\right)
\nn 
\end{eqnarray}
where for $m-n>1$, we have, 
\begin{eqnarray}
{\tilde N}^{m,n} &  =& N^{(m,n)}+\sum_{k= {\rm max}\{0,m-1-d\}}^{m-n-2} (-1)^{k+1}
N^{(k/2,m-2-k,n)}
\label{ntilde}
\end{eqnarray}
Restoring dependence on the fugacities $\{\mu_{i}\}$ we can also expand the
index in characters of the global symmetry group and perform a
similar decomposition for the coefficient of each character.   
The alternating signs in this last expression correspond to potential
cancellations between different (semi-)short multiplets contributing to the
index. Given the value of the index ${\mathcal I}$ as a function of 
$t$ and $y$ it is possible to read off the numbers $N^{(m,m)}$ and 
$\tilde{N}^{(m,n)}$. In general this partial information is only
enough to provide certain lower bounds on the degeneracies. However in
some special cases there is only one non-vanishing term on the RHS 
of (\ref{ntilde}) and we can therefore uniquely determine 
the degeneracy of the corresponding multiplets.  In particular, this
is the case for half-BPS short multiplets $S(m,m)$ $m=0,\ldots d$
but also for semi-shorts $SS(j,d-1,d-1)$. As explained in \cite{DBG},
the former multiplets correspond to the Borel-Moore homology of the
target while the latter are in one to one correspondence with
holomorphic functions on the corresponding complex variety.  
We note that the numbers of $1$/$4$ BPS multiplets 
$S(m,m-1)$ $m=1,\ldots d$ are also uniquely determined. 
We will refer to these representations of $\ospr$ as 
{\em protected representations}. 
\paragraph{}
Our discussion of the superconformal index is so far very similar to
the corresponding discussion in higher-dimensional field
theories. However, in the particular setting of a quantum mechanical
sigma model there are also some new features.  The
Witten index of a particle moving on a Riemannian manifold is
invariant under smooth changes of the metric and is essentially a 
topological invariant of $\mathcal{M}$. It is natural to look for a
similar interpretation of the superconformal index. As in the classic
analysis of \cite{Witten:1982a,Witten:1982b} the starting point is
to reinterpret the supersymmetric vacua as cohomology classes of the
supercharge. Using the mapping of states in the Hilbert space to forms
and working with respect to the preferred complex
structure, the supercharge $s$ acts as, 
\begin{eqnarray}
s = \frac{1}{\sqrt{\mu}}\left(\bar{\partial}+\mu\bar{\partial}K\wedge
\right)
\nonumber 
\end{eqnarray}
where $K$ is the hyper-K\"{a}hler potential and $\bar{\partial}$ is
the Dolbeault operator on $\mathcal{M}$. As the spectrum of
$\mathcal{H}$ is discrete,the usual Hodge theoretic argument implies
that we should identify the space of states contributing to the index
with the cohomology of $s$. The kernel of $s$ consists of forms $\beta$
which can be written as, 
\begin{equation}
\beta=\exp(-\mu K)\, \alpha 
\label{alpha}
\end{equation}
where $\alpha$ is any $\bar{\partial}$-closed form on $\mathcal{M}$:
$\bar{\partial}\alpha=0$. This is well-defined, 
as we have assumed that $K$ is a function on $\mathcal{M}$. More
generally one should consider $s$ acting on sections of an appropriate
line bundle over $\mathcal{M}$. Similarly if $\beta$ is
$\bar{\partial}$-exact with $\beta=\bar{\partial}\gamma$ for some form
$\gamma$ then it is easy to check that $\alpha=s.(\exp(-\mu
K)\,\gamma)$. Hence we may formally identify the cohomology of $s$
with that of the Dolbeault operator $\bar{\partial}$. Importantly 
the presence of the convergence factor $\exp(-\mu K)$ means that the
$L^{2}$-cohomology of $s$ with respect to the inner product 
(\ref{eqn:L2product}) corresponds to the cohomology of
$\bar{\partial}$ acting on forms which are $L^{2}$ with respect to the
modified inner product, 
\begin{eqnarray} 
\left\langle\alpha,\beta\right\rangle_{\mu}  & =  & \int_M \alpha \wedge
*\bar{\beta}\,\exp\left(-2\mu K \right)
\label{eqn:L2product2} \end{eqnarray} 
This is quite different from the standard $L^{2}$-Dolbeault cohomology
of $\mathcal{M}$. For example, in the case of flat space
$\mathcal{M}=\mathbb{C}^{2n}$, the relevant Hilbert space includes all
polynomials in the complex coordinates (as well as 
polynomial-valued forms). 
\paragraph{}
The cohomology $H(\mathcal{M},s)$ described above 
is graded according to holomorphic
degree, 
\begin{eqnarray}
 H\left[\mathcal{M},s\right] & = & 
\oplus_{p,q=0}^{2d}\,  H_{p,q}\left[\mathcal{M},s\right]
\nonumber 
\end{eqnarray}
In the superconformal index, each $(p,q)$-form is weighted with a factor
$y^{p-d}$. Setting the global fugacities to unity the
superconformal index becomes, 
\begin{eqnarray}
\mathcal{I}(y) & = & \frac{1}{y^{d}} \sum_{p, q} (-1)^{p+q} y^{p} {\rm dim}
H_{p,q}\left[\mathcal{M}, s\right]
\nonumber
\end{eqnarray}
which is the analog of Hirzebruch $\chi_{y}$ genus in Dolbeault
cohomology. 
\paragraph{}
The discussion given above is really only applicable for smooth
manifolds and needs to be modified to apply to singular
hyper-K\"{a}hler cones. The approach described in \cite{DBG}, is to
replace the singular cone with its equivariant symplectic resolution
which yields a smooth space. Although this breaks superconformal
invariance, it preserves the two Cartan generators of the $SU(2|1)$
little group corresponding to the index. The cohomology described
above can be reformulated as in terms of sheaf cohomology and the
index can then be understood as an equivariant Euler character of the
resolved space. This approach also yields a concrete formula for the
index as a sum over fixed points of an abelian group action on this
space. We will use this formula below to discuss some simple examples
in Section 4 below.

\paragraph{}
At least formally, the states with $\mathcal{H}=0$ which contribute to
the index are identified with $\bar{\partial}$-cohomology 
classes of $(p,q)$-forms on 
$\mathcal{M}$. The special case $q=0$, which occurs for
eigenstates of the $\mathfrak{usp}(4)$ Cartan generator $\mathbb{M}$ 
with eigenvalue equal to $-d$ corresponds to cohomology classes of
holomorphic forms.  
Interestingly one may check that, for a subset of the (semi-)short $\ospr$ 
representations listed above, {\em all} the $\mathcal{H}=0$ states
correspond to holomorphic forms. We will call these {\em holomorphic
representations}. They are,  
\begin{eqnarray}
S(d,n) &\qquad{}& n\leq d \nonumber \\ 
SS(j,d-1,n) & \qquad{} & n\leq d-1
\label{eqnarray}
\end{eqnarray}
\paragraph{} 
So far we have described the superconformal index in the Hamiltonian
formalism where it corresponds to a particular trace over the spectrum
of superconformal QM. The standard approach to computing indices in
supersymmetric quantum mechanics proceeds by representing the index in
question as a Euclidean functional integral. This naturally requires
us to pass to the Lagrangian formulation. As discussed above the 
superconformal index is 
essentially the Witten index of a quantum mechanical system with
Hamiltonian, 
\begin{eqnarray}
\mathcal{H} & =&
\mathbb{L}_{0}+2\mathbb{J}_{3}+2\mathbb{M} 
\nonumber 
\end{eqnarray}
In terms of the coordinates and their canonical momenta this
corresponds to the classical quantity, 
\begin{align} \label{eqn:EHamiltonian} 
\begin{aligned} \mathcal{H}_{cl} &=
  \frac{1}{2\m}g^{\m\n}\Pi_{\m}\Pi_{\n} +
  \frac{1}{4\m}R_{\m\n\r\s}\adj{\psi}{\m}\adj{\psi}{\n}\psi^{\r}\psi^{\s}
  + \frac{\m}{2}D^{\m}D_{\m} \\ &-
  \mathcal{I}^{\m}_{~\n}D^{\n}\Pi_{\m}+i\adj{\psi}{\m}\psi^{\n}\nabla_{\m}
D^{\mathcal{I}}_{\n}
  + g_{\m\n}\adj{\psi}{\m}\psi^{\n}. \end{aligned} 
\end{align}
where $D^{\mathcal{I}\, \mu}=\mathcal{I}^{\mu}_{\nu}D^{\nu}$. 
We can then obtain a corresponding
Lagrangian by performing a Legendre transform,
\[ \L' = P_{\m}\dot{X}^{\m} - \mathcal{H}_{cl}, \qquad 
\dot{X}^{\m} = \frac{\pd \mathcal{H}_{cl}}{\pd P_{\m}},\]
to find, 
\begin{align} \label{eqn:ELagrangian} \begin{aligned} \L' 
&= \m\left(\frac{1}{2}g_{\m\n}\dot{X}^{\m}\dot{X}^{\n} +
  \omega_{\m\n}^{\mathcal{I}}D^{\n}\dot{X}^{\m}\right) -
\frac{1}{4\m}R_{\m\n\r\s}\adj{\psi}{\m}\adj{\psi}{\n}\psi^{\r}\psi^{\s}
\\ &+ ig_{\m\n}\adj{\psi}{\m}\left( 
\frac{D\psi^{\n}}{Dt} + i\psi^{\n}+\mathcal{I}^{\n}_{~\r}\psi^{\r}\right). 
\end{aligned} \end{align}
where $\omega^{\mathcal{I}}_{\mu\nu}$ is the K\"{a}hler form on $\mathcal{M}$ 
in the complex structure $\mathcal{I}$ picked out by the BPS condition.  
Physically, we have a supersymmetrised coupling of the original
$\s$-model to the magnetic field with vector potential,
\[ A_{\m} = \omega^{\mathcal{I}}_{\m\n}D^{\n}.\]
This model has $\mathcal{N}=(1,1)$ supersymmetry generated by
$\left\{q,s,\mathcal{H}_{cl}\right\}$. A novel feature of this Lagrangian
is that  the little group $\su(2|1)$ is also linearly realised 
as a ``global''  supersymmetry with central extension. See
\cite{AS:thesis} 
for further details.  
\paragraph{}  
A striking feature of the construction given in this section is that
the full spectrum of the theory (not just the index) is independent of
the parameter $\mu$. Changing the value of $\mu$ simply corresponds
to a change of basis for the superconformal algebra. It is natural to
try to exploit this feature to simplify the dynamics. In
the flat space case $\mathcal{M}=\mathbb{C}^{2n}$, 
the eigenstates of the bosonic part ot the operator $\mathcal{H}$ 
are the Landau levels of a particle in a constant background magnetic field
proportional to $\mu$. The fermionic degrees of freedom correspond to
spin degrees of freedom which couple to the magnetic field in the
standard way. Taking the limit $\mu \rightarrow \infty$ isolates the
states in the lowest Landau level. Taking acount of the fermions,
these are precisely the states with $\mathcal{H}=0$ which contribute
to the superconformal index. It is possible to take a similar limit of
the Lagrangian (\ref{eqn:ELagrangian}) in the general
case\footnote{For a more precise treatment of this limit it is
  necessary to resolve the singularity using the approach of \cite{DBG}}.  
After rescaling the fields, this leads
to the first-order Lagrangian,  
\begin{align} \label{eqn:ELagrangian2} \begin{aligned} \L'_{\mu \rightarrow \infty} 
&=  \omega_{\m\n}^{\mathcal{I}}D^{\n}\dot{X}^{\m} + ig_{\m\n}\adj{\psi}{\m} 
\frac{D\psi^{\n}}{Dt}. 
\end{aligned} \end{align}
First order Lagrangians of this type arise in geometric quantization
of symplectic manifolds and it would be interesting to investigate
this approach in the present context.  

\section{Examples and Discussion}

The simplest possible example of a hyper-K\"{a}hler
manifold with a triholomorphic homothety 
is flat $\mathbb{R}^{4}=\mathbb{C}^{2}$. The isometry group is
$H=SO(4)=SU(2)_{L}\times SU(2)_{R}$ of which the factor $SU(2)_{R}$ becomes
an R-symmetry subgroup of $OSp(4^{*}|4)$. The remaining
factor, $SU(2)_{L}$ provides a global symmetry, with Cartan generator 
$L$, which we can use to refine the superconformal index. Hence we
compute, 
\begin{eqnarray}
\mathcal{I}\left[\mathbb{C}^{2}\right] & = & 
{\rm Tr}\left[ (-1)^{F}e^{-\beta \mathcal{H}}\, 
t^{R}y^{N}x^{L}\right] 
\nn 
\end{eqnarray}
For flat space, superconformal quantum mechanics essentially reduces
to decoupled bosonic and fermionic harmonic oscillators and the index
may be computed straightforwardly to give, 
\begin{eqnarray}
\mathcal{I}\left[\mathbb{C}^{2}\right] & = & 
\frac{t}{y}\,  \frac{(1-yx)(1-y/x)}{(1-tx)(1-t/x)} \label{c2}
\end{eqnarray}
States contributing to the index lie in irreducible representations of
the global symmetry. Thus the index can be expanded as, 
\begin{eqnarray}
\mathcal{I}\left[\mathbb{C}^{2}\right] & = & \sum_{l=0}^{\infty} 
\mathcal{I}_{l}(t,y)\chi_{l}(x) 
\label{c3}
\end{eqnarray}
where $\chi_{l}(x)=\sum_{j=0}^{l} x^{l-2j}$ is the character of the
spin $l/2$ representation of $SU(2)_{L}$. Comparing (\ref{c3}) with
(\ref{c2}) we obtain,
\begin{eqnarray} 
\mathcal{I}_{0}  & = & t\left(\chi_{1}(y)-t\right) \nonumber \\ 
\mathcal{I}_{l} & =& t^{l}\left(1+t^{2}-t\chi_{1}(y)\right) \qquad{}
  \qquad{} l\geq 1  \nonumber 
\end{eqnarray}
Although we cannot determine the spectrum of (semi-)short multiplets
from the index alone, in this case the index is saturated by the protected
multiplets described above whose degenercies are uniquely fixed.    
The minimal spectrum consists
of the following direct sum of short and semi-short representations of 
$\osp(4^{*}|4)$ 
each transforming in a particular representation
\footnote{In the following, the integer $l$ appearing as 
the second entry in each square bracket
corresponds to the $SU(2)_{L}$ representation with spin $l/2$.} of the
global symmetry $SU(2)_{L}$;
\begin{equation} 
\left[S(1,1), \, 0\right] \oplus \left[S(1,0), \, 1\right]  \oplus 
 \sum_{l=2}^{\infty} \left[SS\left(\frac{l}{2}-1,0,0\right), l \right]
\end{equation}
In this case it is easy to check directly that this coincides with the
true spectrum of short multiplets. In particular, each of the
$OSp(4^{*}|4)$ representations appearing, in addition to being
protected, is one of the holomorphic 
representations described above. The resulting $E=0$ states are in
$1:1$ correspondence with the holomorphic forms on $\mathbb{C}^{2}$
and the superconformal index coincides with the index of the Dolbeault 
operator acting on polynomial-valued forms.  
\paragraph{} 
The moduli space of 
Yang-Mills instantons on $\mathbb{R}^{4}$ provides an interesting
and highly non-trivial example of superconformal quantum mechanics. Let
$\mathcal{M}_{K,N}$ denote the moduli space of $K$ Yang-Mills
instantons of gauge group $U(N)$. The ADHM construction provides a
description of $\mathcal{M}_{K,N}$ as a quotient of a flat space 
$\mathbb{R}^{4K^{2}+4KN}$. The quotient space inherits three
independent complex structures from those of flat space resulting in a
hyper-K\"{a}hler manifold of quaternionic dimension $d=KN$. The action
of the flat-space dilatation operator also descends 
to provide a triholomorphic homothety on the quotient space. As a
result the quantum mechanical $\sigma$-model with target space
$\mathcal{M}_{K,N}$ has $\osp(4^{*}|4)$ superconformal symmetry.     
\paragraph{}
The instanton moduli space also has an $SU(2)_{L}\times SU(N)$ global
symmetry corresponding to self-dual rotations and global gauge
transformations in Euclidean spacetime. Introducing a fugacity $x$ for 
the Cartan generator $L$ of $SU(2)_{L}$ and fugacities $z_{i}$, with
$i=1,\ldots, N$, for the Cartan generators $J_{i}$ of $SU(N)\subset
U(N)$ we define a refined superconformal index for $\mathcal{M}_{K,N}$
as, 
\begin{eqnarray}
\mathcal{I}_{K,N} & = & 
{\rm Tr}\left[ (-1)^{F}e^{-\beta \mathcal{H}}\, 
t^{\mathbb{T}}y^{\mathbb{N}}x^{L} \prod_{i=1}^{N}\, z_{i}^{J_{i}}\right] 
\nn 
\end{eqnarray}
\paragraph{}
As the ADHM moduli space has singularities corresponding to small
instantons, a regularisation is required to make the index well
defined. This example fits into a large class of examples discussed in
\cite{DBG} which admit an equivariant symplectic resolution. Working on the
resolved space then yields a closed formula for the index as a sum 
over fixed points of the maximal torus of 
$SU(2)_{R} \times SU(2)_{L}\times SU(N)$. The fixed points 
\cite{Nak1999,Nek1,Nek2} are labelled by 
$N$-coloured partitions of the instanton number $K$. Thus for each
fixed point we introduce a vector $\vec{Y}$ of Young tableaux with
components $Y_{i}$, $i=1,2,\ldots,N$,
with a total of $K$ boxes: $||\vec{Y}||:=\sum_{i=1}^{N}|Y_{i}|=K$. We define the
functions of a box, $s=(a,b)$, at row $a$ and column $b$ in the Young
tableau $Y_i$ as, 
\be
f_{ij}(s):=-a_i(s)-l_j(s)-1\,,\quad g_{ij}(s):=-a_i(s)+l_j(s)\,,
\ee
where $a_i(s):=Y_{ia}-b$ the arm length and $l_j(s):=(Y_j^\vee)_b-a$ the leg
length relative to $Y_j$. Using the fixed point formula eqn (1.2) of \cite{DBG},
we can write the superconformal index of
$\mathcal{M}_{K,N}$ as, 
\baa\label{Nekpartfn}
\mathcal
{\mathcal I}_{k,N}
& =\sum_{\vec{Y}:\,\, ||\vec{Y}||=K}\prod_{i,j=1}^N \,\,\prod_{s\in
Y_i}\text{PE}\left[t^{g_{ij}(s)-1}x^{f_{ij}(s)}\frac{z_i}{z_j}(1-t/y)(1-t
y)\right]\,,
\eaa
where the summation is 
over vectors of Young tableaux corresponding to all $N$-coloured
partitions of $K$ and $\text{PE}$ denotes the Plethystic exponential.
\paragraph{}
The resulting index coincides
with the $K$-instanton contribution to the Nekrasov
partition function for a particular supersymmetric gauge theory.   
Specifically, we should identify $\mathcal{I}_{K,N}$ with the
K-instanton contribution, 
$$
\mathcal{Z}_{K}\left[\{a_{i}\},M, \epsilon_{1},\epsilon_{2}\right]
$$ 
to the partition function of a $\mathcal{N}=1$ supersymmetric $SU(N)$ gauge
theory in five dimensions compactified on $\mathbb{R}^{4}\times S^{1}$
and subjected to an $\Omega$-background in the non-compact directions
with deformation parameters $\epsilon_{1}$ and $\epsilon_{2}$. In
addition to an $SU(N)$ vector multiplet of $\mathcal{N}=1$
supersymmetry in five dimensions, the model includes an adjoint
hypermultiplet of mass $M$. As usual the partition function depends on
complex numbers $a_{i}$, $i=1,\ldots,N$, corresponding to VEVs for the
scalar fields in the vector multiplet which parametrise the Coulomb
branch. The dictionary between the parameters of the index and those
of the 5d gauge theory is, 
$$ 
z_{i}=\exp(a_{i}), \qquad{} y=\exp(M),  \qquad{} t=\exp\left(
\frac{\epsilon_{1}+\epsilon_{2}}{2}\right),   \qquad{}  x=\exp\left(
\frac{\epsilon_{1}-\epsilon_{2}}{2}\right),
$$
For this model, our superconformal 
index is the same object studied by Kim et al in
\cite{K3L2} where the identification with the Nekrasov partition
function described above was anticipated. 
\paragraph{}
The simplest non-trivial case is that of a single instanton of gauge
group $U(2)$ and the moduli space is $\mathcal{M}_{1,2}=\mathbb{C}^{2}\times 
({\mathbb{C}^{2}}/{\mathbb{Z}_{2}})$. Removing the overall factors
corresponding to the center of mass, the index of the centered moduli
space is a sum over the contributions of two fixed points,   
\begin{eqnarray}
\mathcal{I}\left(\mathbb{C}^2/\mathbb{Z}_{2}\right) & = & \frac{\left(1-\frac{t}{y}\frac{z_{1}}{z_{2}}\right)\left(1-yt\frac{z_{1}}{z_{2}}\right)}
{\left(1-\frac{z_{1}}{z_{2}}\right)\left(1-t^{2}\frac{z_{1}}{z_{2}}\right)} \,\,+\,\,\frac{\left(1-\frac{t}{y}\frac{z_{2}}{z_{1}}\right)\left(1-yt\frac{z_{2}}{z_{1}}\right)}
{\left(1-\frac{z_{2}}{z_{1}}\right)\left(1-t^{2}\frac{z_{2}}{z_{1}}\right)}
\nonumber 
\end{eqnarray}
which gives, 
\begin{eqnarray} 
\mathcal{I}\left(\mathbb{C}^2/\mathbb{Z}_{2}\right) & = & \frac{1}{\left(1-t^{2}\rho^{2}\right)
\left(1-\frac{t^{2}}{\rho^{2}}\right)}\, 
\left[1+t^{4}-2t^{2}\chi_{2}(\rho)+ t(t^{2}+1)\chi_{1}(y)\right]\nonumber \\ 
\label{I12}
\end{eqnarray}
Here $\rho^{2}=z_{1}/z_{2}$ is the fugacity for the Cartan generator
of the global symmetry, $SU(2)_{G}$, corresponding to large gauge
transformations of the instanton. Expanding the index in $SU(2)_{G}$
characters yields, 
\begin{eqnarray}
\mathcal{I}\left(\mathbb{C}^2/\mathbb{Z}_{2}\right) & = & \sum_{l=0}^{\infty} 
{\mathcal{I}}_{2l}(t,y)\chi_{2l}(\rho) 
\label{d1}
\end{eqnarray}
where, comparing with (\ref{I12}), we obtain,
\begin{eqnarray} 
{\mathcal{I}}_{0}  & = & 1+t\left(\chi_{1}(y)-t\right) \nonumber \\ 
{\mathcal{I}}_{l} & =& t^{2l}\left(1+t^{2}-t\chi_{1}(y)\right) \qquad{}
  \qquad{} l\geq 1  \nonumber 
\end{eqnarray}
As before we can determine the minimal spectrum 
of short and semi-short representations of 
$OSp(4^{*}|4)$ consistent with this result for the index
\footnote{As before the integer appearing as 
the second entry in each square bracket
corresponds to the global $SU(2)$ representation with spin $l/2$.};
\begin{equation} 
\left[S(0,0), \, 0\right] \oplus \left[S(1,1), \, 0\right]  \oplus 
 \sum_{l=2}^{\infty} \left[SS\left(l-1,0,0\right), 2l \right]
\end{equation}
All the representations appearing are of protected type. 
With the exception of the single state $S(0,0)$, the
representations appearing are also holomorphic and their $E=0$ states 
correspond to holomorphic forms on $\mathbb{C}^{2}/\mathbb{Z}_{2}$. 
Indeed the resulting spectrum can be understood by starting from the
minimal spectrum on $\mathbb{C}^{2}$ given above and keeping only those
forms even under the action of $\mathbb{Z}_{2}$ then adding a single 
``extra'' state $S(0,0)$ which corresponds instead to a $(1,1)$-form. 
An extra state of this type 
also arises in other related contexts. In particular, it is also
present in the ordinary $L^{2}$ cohomology of the instanton moduli space. 

\paragraph{}
In the case of arbitrary rank and instanton number, we can use the
results of \cite{DBG} to determine the 
exact spectrum of the special protected multiplets described above.  
In particular  $1$/$2$ BPS multiplets $S(n,n)$, for
$n=0,1,\dots,d=NK$, are all singlets under 
the global symmetry, and the corresponding degeneracies $N^{(n,n)}$ 
are given as, 
\begin{eqnarray}
N^{(n,n)} & = & b_{2(d-n)}
\nonumber
\end{eqnarray}
where the non-negative integer $b_{2r}$ 
is the dimension of the Borel-Moore Homology group of
degree $2r$. These dimensions are the coefficients of the corresponding
Poincare polynomial, 
\begin{eqnarray}
P^{(N,K)}\left(q\right) & = & \sum_{r=0}^{NK}\,b_{2r}
\, q^{2r} \nonumber 
\end{eqnarray}
As shown in \cite{DBG}, the Poincare polynomial appears as a particular 
$t\rightarrow 0$ limit of the superconformal index taken with $q=y/t$
held fixed. Taking this limit in the
fixed-point sum (\ref{Nekpartfn}) yields eqn (4.8) of \cite{DBG}, 
\baa\label{Nekpartfn5}
P^{(N,K)}\left(\frac{y}{t}\right)
& =\sum_{\vec{Y}:\,\, ||\vec{Y}||=K}\prod_{j=1}^N
\,\,\left(\frac{y}{t}\right)^{2N|Y_{j}|-2j\ell(Y_{j})}.
\eaa
Where $|Y_{j}|$ and $\ell(Y_{j})$ are the weight and length of the
partition corresponding to the Young Tableau $Y_{j}$ respectively. 
This agrees with the standard formula for the Poincare polynomial 
due to Nakajima \cite{Nak1999} and yields a closed formula for the
degeneracy $N^{(n,n)}$ of the half BPS multiplet $S(n,n)$ as the
number of N-coloured partitions of total weight $K$ satisfying a particular
linear constraint, 
\bea
N^{(n,n)} & = & \left|\left\{\vec{Y} :\,\, ||\vec{Y}||=K, 
\,\,\sum_{j=1}^{N} j\ell\left(Y_{j}\right)=n \right\}\right| 
\nonumber 
\end{eqnarray}
This set of states was also considered in \cite{ABS:1997} and was 
found to correctly account for the chiral primaries of the $(2,0)$
theory of type $A_{N-1}$ in DLCQ.  
\paragraph{}
The second set of protected states form the multiplets, $SS(r/2,
d-1, d-1)$ for integer $r\geq 2$ (together with $S(d,d-1)$ for $r=1$ and
$S(d,d)$ for $r=0$) which are in one to one correspondence with the 
holomorphic functions 
on the instanton moduli space. For each irreducible representation $R$ of the
global symmetry, the corresponding degeneracy $N^{(j,d-1,d-1)}_{R}$ 
can be extracted from the Hilbert series;  
\begin{eqnarray}
\text{HS}\left(t,x,Z \right) & := & 
\text{Tr}_{H^{0,0}}\left(t^{R}x^{L}\prod_{i}z_{i}\right)
\nonumber 
\end{eqnarray}
Here $H^{0,0}$ denotes the space of holomorphic functions on 
$\mathcal{M}_{K,N}$. More precisely, if we expand the Hilbert series
in characters $\chi_{R}$ of 
the irreducible representations $R$ of the global symmetry group, 
\begin{eqnarray}
 \text{HS}\left(t,x,Z \right) & = & \sum_{R} \, \text{HS}_{R}(t)\,
\chi_{R}\left(x,Z\right) 
\label{HSexpand1}
\end{eqnarray}
then we have,  
\begin{eqnarray}
\text{HS}_{R}\left(t\right) & = & N^{(d,d)}_{R}\,+\, N^{(d,d-1)}_{R} t \,+\, 
\sum_{r=0}^{\infty}\, N^{(r/2,d-1,d-1)}_{R}\, t^{2+r}
\label{HSexpand2}
\end{eqnarray}
The Hilbert series itself is the coefficient of the highest power of
$y$ in the expansion of the superconformal index \cite{DBG} 
and can be extracted by taking a $y\rightarrow \infty$ limit of
the latter. Taking the limit in equation (\ref{Nekpartfn}), we obtain
the explicit formula, 
\baa\label{Nekpartfn2}
 \text{HS}\left(t,x,Z \right)
& =\sum_{\vec{Y}:\,\, ||\vec{Y}||=K}\prod_{i,j=1}^N \,\,\prod_{s\in
Y_i}\text{PE}\left[t^{g_{ij}(s)-1}x^{f_{ij}(s)}\frac{z_i}{z_j}
(1+t^{2})\right]\,,
\eaa
where the sum over $N$-vectors $\vec{Y}$ whose components are Young
Tableaux is the same as in (\ref{Nekpartfn}). This result agrees with the
existing formulae for the Hilbert series of instanton moduli space 
\cite{Ami1}. The degeneracies of each protected multiplet 
are then uniquely determined by comparing 
equation (\ref{Nekpartfn2}) with (\ref{HSexpand1}, \ref{HSexpand2}).  
To make this explicit one needs to expand the fixed point formula for
the Hilbert series in terms of characters of $SU(2)\times SU(N)$ as we
did for the case $k=1$, $N=2$ above.  
\paragraph{}
The authors thank Alec Barns-Graham and Ami Hanany for useful discussions.

\end{document}